\def\nums{\nu_{\rm ms}}

\def\msol{{\rm M}_{\odot}}
\def\nurot{\nu_{\rm rot}}

\def\rms{R_{\rm ms}}
\def\rg{R_{\rm g}}
\def\req{R_{\rm eq}}
\def\qgr{q_{\rm GR}}

\def\citep{\cite}

\documentstyle[twocolumn,prd,aps]{revtex}

\begin{document}
\input epsf
%
% Petites macros
%
\newcommand{\ddp}[2]{\frac{\partial #1}{\partial #2}}
\newcommand{\ddps}[2]{\frac{\partial^2 #1}{\partial #2 ^2}}

\draft
\title{Small strange stars and marginally stable orbit in Newtonian
theory}

\author{Julian Leszek Zdunik}
%\email{jlz@camk.edu.pl}%
%\affiliation{%
\address{%
N. Copernicus Astronomical Center, Polish
           Academy of Sciences, Bartycka 18, PL-00-716 Warszawa,
           Poland,\\ e-mail: jlz@camk.edu.pl
}%
\author{Eric Gourgoulhon}
%\email{Eric.Gourgoulhon@obspm.fr}%
%\affiliation{%
\address{%
D\'epartement d'Astrophysique Relativiste et de Cosmologie
-- UMR 8629 du CNRS, Observatoire de Paris,\\ F-92195 Meudon Cedex,
France, e-mail: Eric.Gourgoulhon@obspm.fr}

\date{15 December 2000}
\maketitle
\begin{abstract}
It is shown that for very rapidly rotating low mass strange stars
the marginally stable orbit is located above the stellar surface.
This effect is explained by the very important role of the
oblateness of the rotating strange star. The comparison with some
``academic'' examples is presented. This feature is purely
Newtonian in its nature
and has nothing to do with relativistic marginally stable orbit.
The effect is very large and cannot be treated in a perturbative way.
It seems that strange stars as a very dense self-bound objects are
the only possibility in Nature to represent these toy models.

%\keywords{dense matter -- equation of state -- stars:}

\end{abstract}


%\maketitle
%
%\section{Introduction}
%%%%%%%%%%%%%%%%%%%%%%%%%%%%%%%%%%
%

Recently marginally stable orbits have been extensively studied as
a crucial input of the most popular model explaining kilohertz
quasi-periodic oscillations (QPOs) in X-ray binaries
(see
\cite{2000ARA&A..38..9005V}
%der Klis 2000
for a review).
Besides neutron stars as a possible source of these QPOs
hypothetical strange stars have been considered.
Strange stars, objects built
of the mixture of u, d, s quarks, in contrast to neutron stars
have no minimum mass; their size extends continuously from 
that of large nuclei (strangelets)
up to the maximum mass configurations of about 2 solar masses
(for a recent review, see \cite{1999hdmh.conf..162M}).
It has been showed
\citep{1999A&A...352L.116S,2000A&A...356..612Z}
%(Stergioulas et al. 1999, Zdunik et al. 2000)
that the marginally stable orbit is located above
the surface of the strange star for relatively high mass even at
very high rotation rates, which is the main difference with
respect to neutron stars \citep{1998ApJ...509..793M}.
%(Miller et al. 1998).
It was suggested \cite{2000A&A...356..612Z}
that, for configurations close to the Keplerian one, defined
as a mass-shedding limit at the equator, the very
large deformation of the strange star leads to the significant
increase of the radius of the marginally stable orbit, resulting
in a gap between this orbit and the stellar surface. Of course in
these considerations, performed for relatively massive ($M>\msol$)
stars, the GR effects seem to play a dominant role. General relativity
predicts that the radius of
the marginally stable orbit for nonrotating star of mass $M$
 is equal to $\approx 8.8 M/\msol$~km, which
is comparable to the radius of a strange star. It is well known
that quadrupole moment of the star could also destabilize the
orbital motion \cite{1998PhysRev...58..10041S},
%(Shibata \& Sazaki 1998),
but it seemed that this
effect is not very large in real physical situations,
at least as far as neutron stars are concerned.
We show that for objects such as self-bound strange stars the
deformations due to the rotation can be so large that the change
in Newtonian gravitational potential could result in the existence
of a marginally stable orbit, even when GR corrections are
completely negligible.

%\section{Results}

The existence of a marginally stable orbit corresponds to the
conditions ${\rm d}V/{\rm d}r=0$ and ${\rm d}^2V/{\rm d}r^2=0$
where $V(r)$ is the effective potential (for the radial part of
motion) for a particle orbiting the central body at a fixed
angular momentum per unit mass $l$. The resulting equations
are discussed in standard textbooks in the case of the Schwarzschild geometry
(see e.g. \cite{1973grav.book.....M}).
%Misner, Thorne \& Wheeler 1973,
%p.~660).
In first approximation, the source of the marginally
stable orbit is the term in $V$ proportional to $1/r^3$ with the
same sign as the gravitational potential (i.e. opposite sign to
centrifugal force). In general relativity it is a natural result of
constants of geodesic motion on circular orbits.

In Newtonian theory this term is generated by the quadrupole moment
of the star $Q$; in the equatorial  plane
the gravitational force per unit mass of the orbiting particle is
\begin{equation}
f_g= -{GM\over r^2}-{3Q\over 2r^4} \ .
\label{fnew}
\end{equation}
With the above convention $Q>0$ for an oblate star and $Q<0$ for a prolate one.
The relativistic definition of $Q$ can be found in
Ref.~\cite{1994A&A...291..155S}.
%Salgado et al. 1994

One can take into account the effect of the oblateness and of the
angular momentum of the
star using standard methods, i.e. via
expansions of the metric functions in powers of
$1/r^2$. In the framework of GR this leads to the following formula
for marginally stable orbit:
\begin{equation}
\begin{array}{lcl}
%\displaystyle{\rms(\qgr,j)\over 3\rg}& =&
\displaystyle{\rms\over 3\rg}& =&
{1\over 2} ({1+\sqrt{1+\qgr/6}})-\\
&&\left({2\over3}\right)^{3/2} j
\sqrt{1+{9\over16}(\sqrt{1+\qgr/6}-1)} \ ,
\end{array}
\label{rmsgjgr}
\end{equation}
where $\qgr=Qc^4/G^2M^3$, $j=Jc/GM^2$, $\rg=2GM/c^2$, $J$ being the total
angular momentum and
we linearized the dependence in $j$. In the case of $Q=0$ Eq.~(\ref{rmsgjgr}) reduces to
the formula derived by Kluzniak and Wagoner \cite{1985ApJ...297..548K}.

In nonrelativistic limit we have
\begin{equation}
\rms\to {3\over2}\Big(\rg+\sqrt{\rg^2+{2Q\over 3M}}\Big)\to\sqrt{3Q\over 2M} \ .
\label{rmsq}
\end{equation}
%\subsection{Models}

In Newtonian theory the marginally stable orbit
can be determined exactly for some
toy models. We will present here results for one- and two-dimensional cases.

In one dimension we consider a bar with given constant linear
density $\rho_l$, i.e. mass of the bar is $M=2 \rho_l R$, where
$R$ is the ``radius'' of the bar equal to the half of its length.
The gravitational attraction on the line which coincides with the
bar at the distance $r$ from the bar center is equal to
$f_g=-GM/(r^2-R^2)$. Of course we have to assume that this field
does not change during orbital motion of the particle, i.e. the
bar rotates with the same angular velocity as the orbiting particle.
The radius of the
marginally stable orbit is then
\begin{equation}
\rms^{\rm bar}=\sqrt{3} R \simeq 1.73 R \ .
\label{rmsbar}
\end{equation}

%%%%%%%%%%%%%%%%%fig1%%%%%%%%%%%%%%%%%%%%%%%%%%%%%%%%%%%%%%%%%%%%%%
%%%%%%%%%%%%%%%%%fig1%%%%%%%%%%%%%%%%%%%%%%%%%%%%%%%%%%%%%%%%%%%%%%
\begin{figure}
\epsfxsize=\hsize
\epsffile{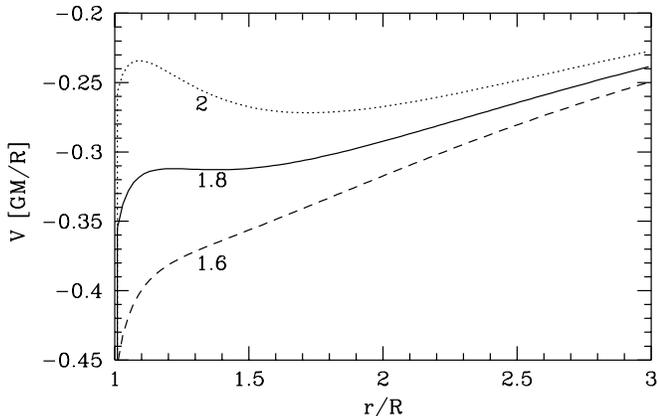}
\caption{Effective potential
of the particle on the circular orbit around the disk when the
square of angular momentum $l^2$ is equal to 1.6, 1.8 and $2\, GMR$.
The solid line corresponds to the case of marginally stable orbit
(the exact value is $l_{\rm ms}^2=1.784 \, GMR$).}
 \label{vdisk}
\end{figure}
%%%%%%%%%%%%%%%%%%%%%%%%%%%%%%%%%%%%%%%%%%%%%%%%%
In two dimensions the corresponding body is an infinitesimally
thin disk of constant surface density $\rho_s$ with mass $M=\pi
\rho_s R^2$. The gravitational force in the plane of the disk at
the distance $r$ is then given by the hypergeometric function $^2F_1$
\begin{equation}
f_g=-{GM\over r^2}{1\over(1+R/r)^{3/2}}\,
{}^2F_1\Big({3\over4},{5\over4},2;q^2\Big) \ ,
\label{fgdisk}
\end{equation}
where $q=2rR/(r^2+R^2)$.
The form of the effective potential [the integral of Eq.~(\ref{fgdisk})
plus a centrifugal term] results
in  the marginally stable orbit
located at the radius
\begin{equation}
\rms^{\rm disk}\simeq 1.2885 R
\label{rmsdisk}
\end{equation}

The angular momentum of the particle at this orbit is then
equal to $l\simeq \sqrt{1.784 GMR}$ (per particle mass).

It should be mentioned that in these examples
expansions of the gravitational potential in powers of $({R/
r})^2$ would be completely unjustified. For example in the case of a
disk the expansion of Eq. (\ref{fgdisk}) leads to the formula:
\begin{equation}
f_g=-{GM\over r^2}{(1+{3\over 8}(R/r)^2+{15\over 64}(R/r)^4 + \dots)} \ .
\label{fgdiskex}
\end{equation}
Using only the second term (quadrupole order) we obtain  the radius
of marginally stable orbit
\begin{equation}
\rms^{\rm disk}[Q] = \sqrt{0.375} R \simeq 0.61 R
\label{rmsdiskQ}
\end{equation}
i.e more than twice smaller than the exact value.

%\subsection{Rotating strange stars}

%%%%%%%%%%%%%%%%%fig2%%%%%%%%%%%%%%%%%%%%%%%%%%%%%%%%%%%%%%%%%%%%%%
%%%%%%%%%%%%%%%%%fig2%%%%%%%%%%%%%%%%%%%%%%%%%%%%%%%%%%%%%%%%%%%%%%
\begin{figure}
\epsfxsize=\hsize
\epsffile{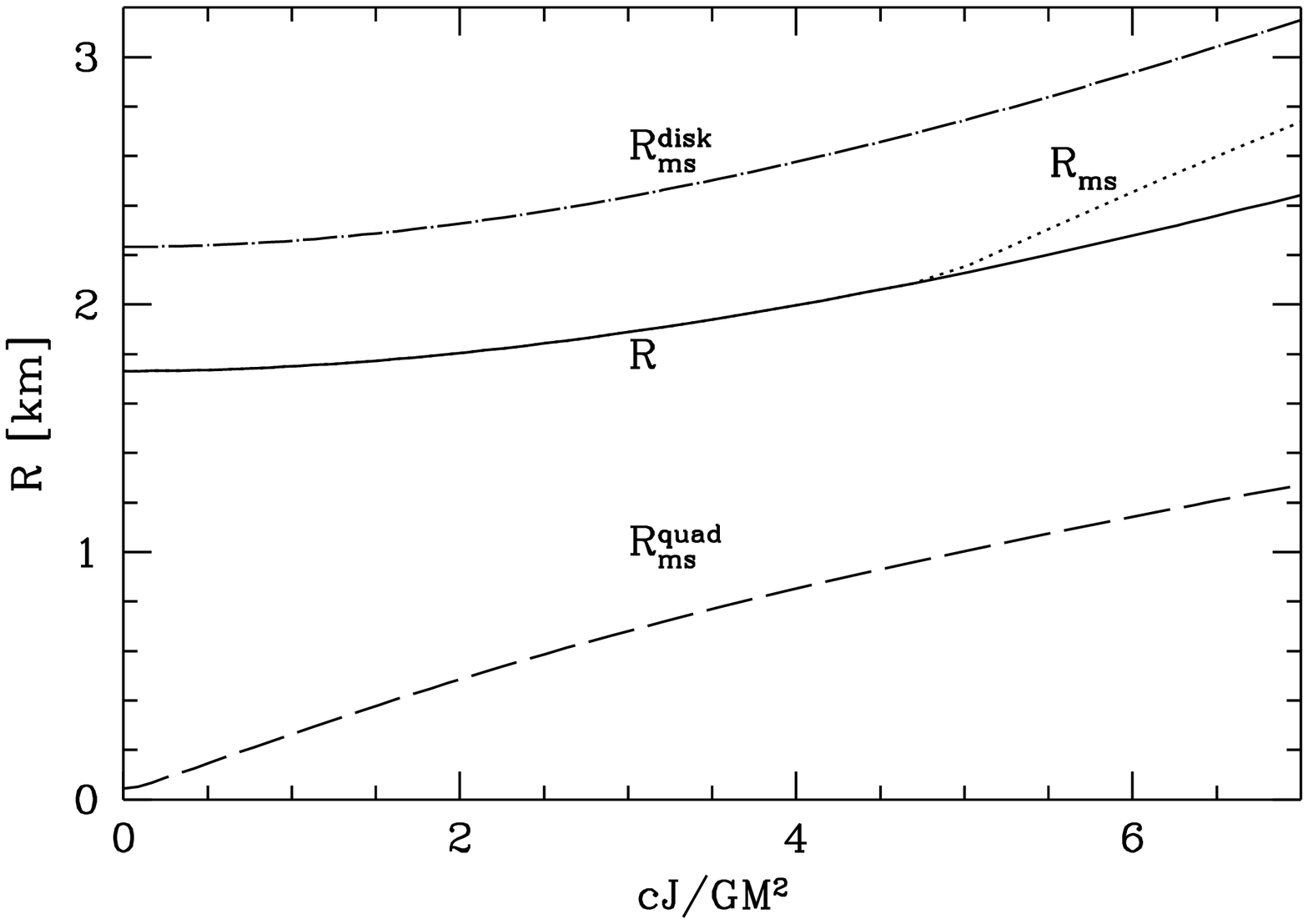}
\caption{Equatorial radius of the strange star $R$ (solid),
 radius of the marginally
stable orbit $\rms$ (dotted), quadrupole approximation of $\rms$ -
$\rms^{\rm quad}$ (dashed), and radius of the marginally stable
orbit if the rotating object would be a disk of the same radius
$\rms^{\rm disk}$. The star has a mass $0.005 \msol$. The value of
$\rms^{\rm quad}$ at $J=0$ corresponds to the relativistic radius
of the marginally stable orbit for nonrotating star which in this
case is equal to 0.044 km. The relativistic term due to the
dependence on $j$ (Eq. \ref{rmsgjgr}) is negligible.}
 \label{rvsj}
\end{figure}
%%%%%%%%%%%%%%%%%%%%%%%%%%%%%%%%%%%%%%%%%%%%%%%%%
In previous papers
%\cite{1999A&A...352L.116S, 2000A&A...356..612Z}
\cite{1999A&A...352L.116S,2000A&A...356..612Z}
%(Stergioulas et al. 1999, Zdunik et al. 2000)
the marginally stable orbit
for rotating strange stars have been determined in the framework of
General Relativity and the authors concluded
that the main difference with respect to neutron stars is the
gap between this orbit and the stellar surface for high rotation rates.

Rotating strange stars are very oblate for large rotation rates
and the Keplerian configurations for a sequence of stars with given
baryon number is achieved not due to the increase of the rotation
frequency $\nurot$,
but due to the deformation of the star
\citep{2000A&A...356..612Z}.
%(Zdunik et al 2000).
It was suggested that this large oblateness is the true reason
which cause the radius of the marginally stable orbit
to be above the stellar surface.

In this paper we try to prove this statement by considering strange stars
for which relativistic effects are very small.
Thus we will focus on the results for small strange stars
where the oblateness
is crucial for the properties of these objects and in particular, is the
source of the marginally stable orbit. The effects
of General Relativity, which are very important in massive dense stars, are
here negligible.
However presented results has been obtained in the framework of GR
using the numerical code developed by Gourgoulhon et al.
\cite{1999A&A...349..851G}.
%Gourgoulhon et al. (1999).

We calculate rotating bare strange stars (without crust) using
the strange matter model
considered as a basic one in previous papers (and called SQM1
in \citep{2000A&A...356..612Z}).

In Fig. \ref{rvsj} we present the equatorial radius of the star $\req$ with
fixed baryon
mass $M=0.005\, \msol$ and the radius of marginally stable orbit. 
We choose such a small object in order to extract Newtonian effects - which
for massive stars are contaminated (and for slowly rotating stars dominated)
by relativistic ones.
We see that
for high rotation rates there exist a gap between stellar surface and the
marginally stable orbit. The curve $R_{\rm ms}^{\rm disk}$ is
defined by Eq.~(\ref{rmsdisk}) with $R=R_{\rm eq}$.
The curve $R_{\rm ms}^{\rm quad}$ corresponds to the quadrupole approximation of
$\rms$ given
by Eq.~(\ref{rmsgjgr}) with  $Q$ calculated in
the framework of GR [Eq.~(7) of
\cite{1994A&A...291..155S}].
%Salgado et al. (1994)].
We see that this quadrupole approximation is wrong and
as in the case of disk the resulting radius is about twice smaller than the
true  value. This conclusion confirms the results of Shibata and Sasaki
\cite{1998PhysRev...58..10041S},
who found necessary to include higher multipole moments in the marginally
stable orbit determination in the case of polytropic EOS.
They concluded that the quadrupole moment is as important as the
spin.
In our model the effect of the rotational deformation of the star is
larger because the matter is self-bound at very high density and the stars
are more oblate. Due to these two reasons for fast rotating strange stars
we are ``closer'' to the disk approximation than to the expansion up to the
quadrupole term (see  Fig. \ref{rvsj}). The value of the quadrupole
moment for small strange stars close to the Keplerian configuration
is $Q_{\rm max}=0.17 MR^2$ which is quite close to the ``disk'' value --
$0.25  MR^2$.

%%%%%%%%%%%%%%%%%fig3%%%%%%%%%%%%%%%%%%%%%%%%%%%%%%%%%%%%%%%%%%%%%%
%%%%%%%%%%%%%%%%%fig3%%%%%%%%%%%%%%%%%%%%%%%%%%%%%%%%%%%%%%%%%%%%%%
\begin{figure}
\epsfxsize=\hsize
\epsffile{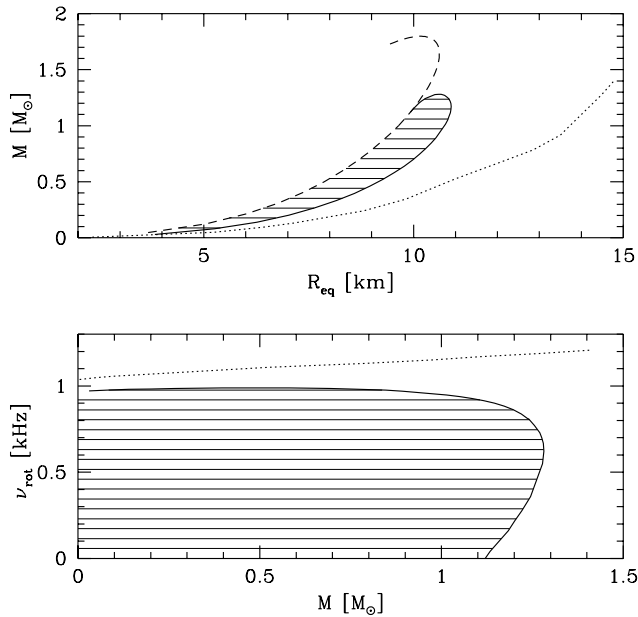}
 \caption{The regions where
equatorial radius of the marginally stable orbit is smaller than
stellar radius (hatched). Solid curve is defined by the equality
$\req=\rms$. The condition $\req<\rms$ is fulfilled between solid
and dotted curve (corresponding to the Keplerian frequency). The
dashed curve represents nonrotating configurations (top panel). In
hatched area the ISCO is defined by the stellar surface. The
region for $1.12 < M/\msol < 1.28 $ corresponds to the
configurations for which for given mass  $\rms$ decreases with
$\nurot$ due to dependence on $j$, then crosses the radius and for
higher $\nurot$ reappears.}
 \label{vmrms}
\end{figure}
%%%%%%%%%%%%%%%%%%%%%%%%%%%%%%%%%%%%%%%%%%%%%%%%%
The behavior of $\req$ and $\rms$ for a star rotating faster and
faster, presented in Fig. \ref{rvsj}, is typical for small mass
stars self-bound at high density. For higher masses (larger than
$1.12\,\msol$ in this model) the radius of marginally stable orbit
is larger than $R$ at $\nurot=0$ and for low rotation rates
decreases due to the relativistic effects and then increases due
to the oblateness (see \citep{2000A&A...356..612Z}
%Zdunik et al. 2000
for figures and discussion).
Of course the point at which $\rms(\nurot)$
gets larger than the equatorial radius $\req$ depend on the mass of the star under
consideration.

In the next figures we present the regions of stellar parameters and
rotation rates for which marginally stable orbit is located
above the stellar surface.

In Fig. \ref{vmrms} we show the boundaries resulting from the
Keplerian frequency and from the condition $\req=\rms$.
We see that in $M-R$ plane the region with
$\rms<\req$ looks rather small
and even for
low-mass stars there exist relatively large area in which $\req<\rms$.
It is also clearly visible that the stars with masses between $1.12$ and
$1.28\,\msol$ enter the region with $\rms<\req$ during slowing down
or acceleration although for high and low rotation rates  $\rms>\req$.
These stars for fixed baryon mass are represented by nearly
horizontal (vertical) lines in top (bottom) panel of Fig. \ref{vmrms}
(``nearly'' because gravitational mass increases
a little due to the rotation).

%%%%%%%%%%%%%%%%%fig4%%%%%%%%%%%%%%%%%%%%%%%%%%%%%%%%%%%%%%%%%%%%%%
%%%%%%%%%%%%%%%%%fig4%%%%%%%%%%%%%%%%%%%%%%%%%%%%%%%%%%%%%%%%%%%%%%
\begin{figure}
\epsfxsize=\hsize
\epsffile{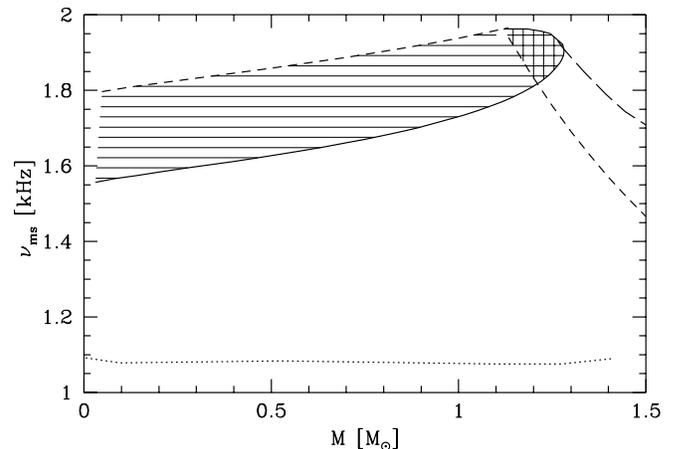}
\caption{The region in the $\nums$ vs $M$ plane where the radius
of the marginally stable orbit is smaller than the stellar radius
(hatched area). In the double hatched region there exist
configurations with $\req<\rms$ (slowly rotating) as well as
$\req<\rms$ (see text). The solid curve is defined by the
condition  $\req=\rms$, the dashed line is the frequency at the
ISCO for nonrotating star, which for $M<1.12\msol$ coincides with
stellar radius and for higher masses is proportional to $1/M$. The
long dashed curve is the maximum $\nums$ for higher mass strange
stars when $\req<\rms$ everywhere.
 The dotted curve is the minimum value
of $\nums$ achieved for Keplerian configurations. }
 \label{fms}
\end{figure}
%%%%%%%%%%%%%%%%%%%%%%%%%%%%%%%%%%%%%%%%%%%%%%%%%

In Fig \ref{fms} we present limits on the frequency of marginally stable
orbit taking into account the conditions $\req<\rms$ which in principle
can be supported by the observations of QPOs
\citep{1999ApJ...520L..37K}.
%(Kaaret et al. 1999).

There exists a
quite large region in the $\nums$-$M$ plane, where this condition is
fulfilled. It should be mentioned that because $\nums$ in non-monotonic function
of $J$ for a star with given baryon mass, one region in $\nums$-$M$
plane is not uniquely defined. Namely
in double hatched area there exist stellar
configurations with $\req<\rms$ (slowly rotating) as well as stars for which
$\req>\rms$. For nonrotating stars we have here $\req<\rms$ but if such a star
rotates faster and faster it moves upwards, turns back at the
solid line and moves downwards having $\req>\rms$ as long as it do not
leave the region between solid lines.
One has similar situation (non-uniqueness)
in the case of high mass stars ($M > 1.28 \msol$), which turn
back at long-dashed line but the condition $\req<\rms$ is fulfilled all the
time. In non hatched region all stars have $\req<\rms$.

%%%%%%%%%%%%%%%%%fig5%%%%%%%%%%%%%%%%%%%%%%%%%%%%%%%%%%%%%%%%%%%%%%
%%%%%%%%%%%%%%%%%fig5%%%%%%%%%%%%%%%%%%%%%%%%%%%%%%%%%%%%%%%%%%%%%%
\begin{figure}
\epsfxsize=\hsize
\epsffile{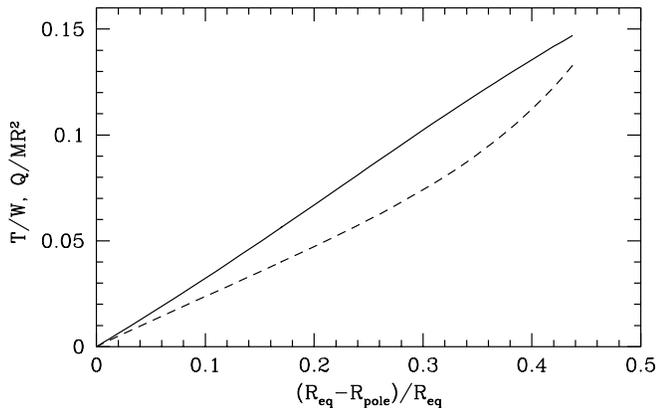}
\caption{The ratio $T/W$ (solid) and the quadrupole moment $Q$
(dashed) as a function of the oblateness of the star along the
sequence of rotating strange stars for which $\req=\rms$ (solid
curves in Figs \ref{vmrms}). Point $(0,0)$ describes nonrotating
configuration. The last points (maxima) correspond to the smallest
stars. The Keplerian limits are located far above and to the right
of this graph ($T/W\sim 0.25$, oblateness $\sim 0.63$) }
 \label{twqob}
\end{figure}
%%%%%%%%%%%%%%%%%%%%%%%%%%%%%%%%%%%%%%%%%%%%%%%%%
From  Fig. \ref{twqob} we see that the oblateness of the star, the
quadrupole moment $Q$ and the ratio of the kinetic to the
potential energy $T/W$ for the rotating stellar configurations
defined by the equality $\req=\rms$ are highly correlated and
almost linearly dependent. Although the frequency at which this
condition is fulfilled is approximately constant for $M<1\,\msol$
(0.9-1 kHz, Fig \ref{vmrms}) the stars with lower mass are much
more oblate at this rotation rate, having proportionally larger
$T/W$ and $Q$. For these configurations large value of $\rms$
results entirely from oblateness  of the star.
Albeit the role of the oblateness for the location of the marginally
stable orbit is most pronouncing in the case of small strange stars
close to the Keplerian velocity, also for massive stars rotating with 
relatively high frequency the deformation of the star is crucial for 
the existence of the marginally stable orbit.

%\section{Discussion and conclusion}

In this paper we showed that the gap between innermost stable
orbit and the surface of the star can be produced by the
oblateness of the star in the framework of Newtonian theory. Due
to the compactness and very high surface density strange stars are
natural candidates for this effect to be a real possibility in
Nature. This is due to two reasons. First: the stars are very
oblate, and the  gravitational force is far from the attraction by
a point mass. Second: the stars are self-bound at very high
density and in practice an oblate small strange star is quite
close to a constant density disk. In contrast to bare strange stars,
%considered in this paper,
for oblate neutron
stars (and also for strange stars with crust) the low-density
region is relatively large making the radius of the star larger
than $\rms$ (or in other words the radius of corresponding
``disk''
is much smaller than the stellar radius).

\begin{acknowledgments}
%\medskip
We are very grateful to P.~Haensel for careful reading of manuscript and helpful
comments and suggestions.
 This research
was partially supported by the KBN grants No. 2P03D.014.13.
The numerical calculations have been performed on computers purchased
thanks to a special grant from the SPM and SDU departments of
CNRS. Main results reported in the present note were obtained during  visit
of one of the authors (J.L.Z) at DARC, Observatoire de Paris-
Meudon,  in July 2000,
within the framework of the  CNRS/PAN  program  Jumelage Astrophysique.
\end{acknowledgments}

\newcommand{\apjl}{Astrophys. J. Lett.}
\newcommand{\aap}{Astron. Astrophys.}
\newcommand{\araa}{Annu. Rev. Astron. Astrophys.}

%\bibliography{oblat}

\end{document}